\documentclass[aps,pra,twocolumn,superscriptaddress]{revtex4}
\usepackage{graphicx}
\usepackage{dcolumn}
\usepackage{color}
\usepackage{bm}
\usepackage{amssymb}
\usepackage{amsmath}
\begin{document}
\title{Spectral characteristics of a modified inverted-Y system beyond rotating wave approximation}
\author{Charu Mishra}
\email[E-mail: ]{charumishra@rrcat.gov.in}
\author{A. Chakraborty}
\author{S. R. Mishra}
\affiliation{Laser Physics Applications Section, Raja Ramanna Centre for Advanced Technology, Indore-452013, India.}
\affiliation{Homi Bhabha National Institute, Training School Complex, Anushakti Nagar, Mumbai-400094, India.}

\begin{abstract}

The spectral properties of a multilevel atomic system interacting with multiple electromagnetic fields, a modified inverted-Y system, have been theoretically investigated. In this study, a numerical matrix propagation method has been employed to study the spectral characteristics beyond the validity regime of the rotating wave approximations. The studied atomic system, comprising of several basic sub-systems \textit{i.e.} lambda, ladder, vee, N and inverted-Y, is useful to study the interdependence among these basic sub-systems. The key features of the obtained probe spectra as a function of coupling strength and detuning of the associated electromagnetic fields show inter-conversion, splitting and shifting of the transparency and absorption peaks. The dressed and doubly dressed state formalism have been utilized to explain the numerically obtained results. This study has application in design of novel optical devices capable of multi-channel optical communication along with switching.
\end{abstract}

\maketitle

\section{Introduction}

Studies on atomic systems interacting with electromagnetic fields has a long and prosperous history. However, in the recent past, multi-level atomic systems have been investigated rigorously either to gain an insight into the fundamental features of the atom-field interactions or to implement the physical behaviors in optical devices. A plethora of cutting edge technologies has already been attributed to the optical phenomena discovered due to the advancement of spectroscopy techniques. The three level atomic systems which facilitated the first observation of quantum interference effects in atomic systems are one of the most basic building blocks for the optical switching devices using Electromagnetically induced transparency (EIT) \cite{Harris:1990, Harris:1st} and electromagnetically induced absorption (EIA) \cite{akulshin:1998}. Apart from these three level systems, namely  $\Lambda$ \cite{Li:4:1995, Li:6:1995,Charu:65:2018}, ladder \cite{MOSELEY:1995, Anil:2009, Dinh:2016} and vee \cite{Zhao:2002, Kang:2014} atomic systems, there are atomic systems consisting higher number of atomic levels which have already been proven to display wonderful physical properties which not only have enriched physics but the application value is enormous too. Two promising four level atomic systems, \textit{i.e.} inverted Y \cite{Gao:2000, Yan:64:2001, Qi:2009, Dong:2012, Sabir:2016, Kavita:2017} and N configuration \cite{Goren:2004, Kong:2007, Abi:2011, Somia:2013, Phillips:2013, Kang:2015, Islam:2017, Tuan:2018}, have already been earmarked for their application in non-linear spectroscopy for producing large Kerr non-linearities in an optical medium \cite{Kou:2010, Yang:2015}. These third order non-linear effects relies on large field strengths to obtain considerably enhanced third order optical susceptibility. Historically, increasing the number of atomic levels interacting with increased number of externally applied electromagnetic fields increases the diversity of observed physical phenomena but also the complexity to a many fold both theoretically and experimentally. 

In this article, a modified inverted-Y system, hereafter denoted as $IY^+$ system, is considered which comprises of all the three basic three-level sub-systems ($\Lambda$, ladder and Vee) and two basic four-level sub-systems (inverted-Y and N). This system is investigated with variation in external field parameters like field strength and detuning. The aim of this study is to investigate this atomic system for cold atoms to control the probe absorption in a desired manner for device applications. Here, we have used a numerical matrix propagation method in which a complete density matrix has been propagated through time in order to obtain the transient characteristics as well as the steady state condition for a probe field absorption. The rotating wave approximations (RWA) are routinely employed in the spectroscopy for finding pertinent informations regarding the interaction of atomic systems with electromagnetic fields in steady-state. Though extremely successful, RWA suffers severely when the atom-field interactions are far from the resonance condition and the applied electromagnetic fields are strong enough to be treated perturbatively. The numerical approach utilized in this article, to solve the Liouville equation is described in detail in \cite{masayoshi:1994, masayoshi:1995} and a complex algebraic form of this method is numerically implemented in this article. The obtained results employing this approach, have also been explained using dressed and doubly dressed state formalism. 

The article is organized as follows. In section \ref{sec:NMP}, the numerical matrix propagation (NMP) technique is discussed. In section \ref{sec:transient}, the study of transient characteristics of inverted-Y system and equivalence of NMP method with the conventional RWA method is established. In section \ref{sec:result}, the probe absorption in $IY^+$ system is explored in steady state regime and the obtained results are discussed. The conclusion of work is presented in section \ref{sec:conc}.

\section{Numerical Matrix Propagation}
\label{sec:NMP}
An in depth description of the numerical solution technique employed in this article has been provided in \cite{masayoshi:1994,masayoshi:1995}. A brief outline of the theoretical background and numerical implementation is provided here for completeness. The generalized electromagnetic radiation composed of M individual field components can be written as,
\begin{equation}
\mathcal{E}(t)=\sum_{p}^{M}\varepsilon_p(\omega_p)\cos(\omega_pt),
\end{equation}
where $\varepsilon_p$ and $\omega_p$ represent the field strength and angular frequency for $p^{th}$ field component.
The Hamiltonian of the composite atom-field system can be written as,
\begin{equation}
H(t)=H_0+H_{I}(t),
\end{equation}
where $H_0$ is the atomic Hamiltonian described in the atomic basis states $\{|\alpha\rangle\}$ as,
\begin{equation}
H_0|\alpha\rangle=\epsilon_i|\alpha\rangle.
\end{equation}
The electromagnetic interaction Hamiltonian $H_I$ can not be written exactly in the atomic basis states due to the infinite sequence of multipole interactions that can originate from the interaction of the atom with the electromagnetic fields. Under the long wavelength approximation (\textit{i.e.} dipole approximation), the interaction can be written as,
\begin{equation}
H_I(t)=-\mu.\mathcal{E}(t).
\end{equation} 
The evolution of the density matrix ($\rho=\sum_{\alpha}|\alpha\rangle\langle\alpha|$) for all the atomic levels can be described by the Liouville equation as,
\begin{equation}\label{eq:liouville}
\frac{\partial}{\partial t}\rho(t)=-\frac{i}{\hbar}\left[H(t),\rho(t)\right]-\frac{1}{\hbar}[R,\rho(t)].
\end{equation}
The first part in right-hand side of the Liouville equation describes the unitary evolution whereas the second part describes the decay of the composite atom-field system. The second part can be written explicitly in terms of the individual decay rates between different states as,
\begin{equation}
\left[R,\rho(t)\right]_{\alpha\alpha}=-\Gamma_{\alpha\alpha}+\sum_{\beta\neq\alpha}\gamma_{\beta\alpha}\rho_{\beta\beta}
\end{equation}
and,
\begin{equation}
\left[R,\rho(t)\right]_{\alpha\beta}=-\Gamma_{\alpha\beta}\ (\alpha\neq\beta)
\end{equation}
The decay rates can also be written as,
\begin{equation}
\Gamma_{\alpha\beta}=\frac{1}{2}\left(\Gamma_{\alpha\alpha}+\Gamma_{\beta\beta}\right)+\Gamma'_{\alpha\beta}
\end{equation}
with the property
\begin{equation}
\Gamma_{\alpha\beta}=\Gamma_{\beta\alpha}
\end{equation}
and 
\begin{equation}
\Gamma_{\alpha\alpha}=\sum_{\beta\neq\alpha}^{N}\gamma_{\alpha\beta},
\end{equation}
where $\gamma_{\alpha\beta}$ are the feeding parameters and $\Gamma'_{\alpha\beta}$ is dephasing factor due to phase changing collisions which is considered zero in this study. To solve Liouville equation \ref{eq:liouville}, density matrix elements are divided into the diagonal and off-diagonal elements denoted by complex variables $\rho_{ii}$ and $\rho_{ij}$.

The evolution of population of state $i$ can be obtained by solving following equation,

\begin{equation}
\begin{split}
\Re(\dot{\rho_{ii}})&=\frac{1}{\hbar}\sum_{p}^{M}\sum_{k}^{N}(\Im(\rho_{ik})\mu_{ki}-\mu_{ik}\Im(\rho_{ki}))\varepsilon_p\cos(\omega_pt)\\&+\sum_{k\neq i}^{N}\gamma_{ki}\Im(\rho_{kk})-\Gamma_{ii}\Re(\rho_{ii}).
\end{split}
\end{equation}
and the equations from which time dependent off-diagonal density matrix elements can be evaluated, are written as,

\begin{equation}
\begin{split}
\Re(\dot{\rho_{ij}})&=\frac{\epsilon_{ij}}{\hbar}\Im(\rho_{ij})\\&+\frac{1}{\hbar}\sum_{p}^{M}\sum_{k}^{N}(\Im(\rho_{ik})\mu_{kj}-\mu_{ik}\Im(\rho_{kj}))\varepsilon_p\cos(\omega_pt)\\&+\Gamma_{ij}\Re(\rho_{ij})
\end{split}
\end{equation}
and
\begin{equation}
\begin{split}
\Im(\dot{\rho_{ij}})&=\frac{\epsilon_{ij}}{\hbar}\Re(\rho_{ij})\\&+\frac{1}{\hbar}\sum_{p}^{M}\sum_{k}^{N}(\mu_{ik}\Re(\rho_{kj})-\Re(\rho_{ik})\mu_{kj})\varepsilon_p\cos(\omega_pt)\\&+\Gamma_{ij}\Re(\rho_{ij}).
\end{split}
\end{equation}
The resulting time series of the propagated matrix $\rho$ can be utilized to obtain the populations of individual states $\Re(\rho_{ii})$ or the coherence of individual transitions $\Im(\rho_{ij})$ which gives absorption between states $i$ and $j$.

\section{Transient characteristics and Comparison with RWA}
\label{sec:transient}

\begin{figure}
\includegraphics[width=8.5 cm]{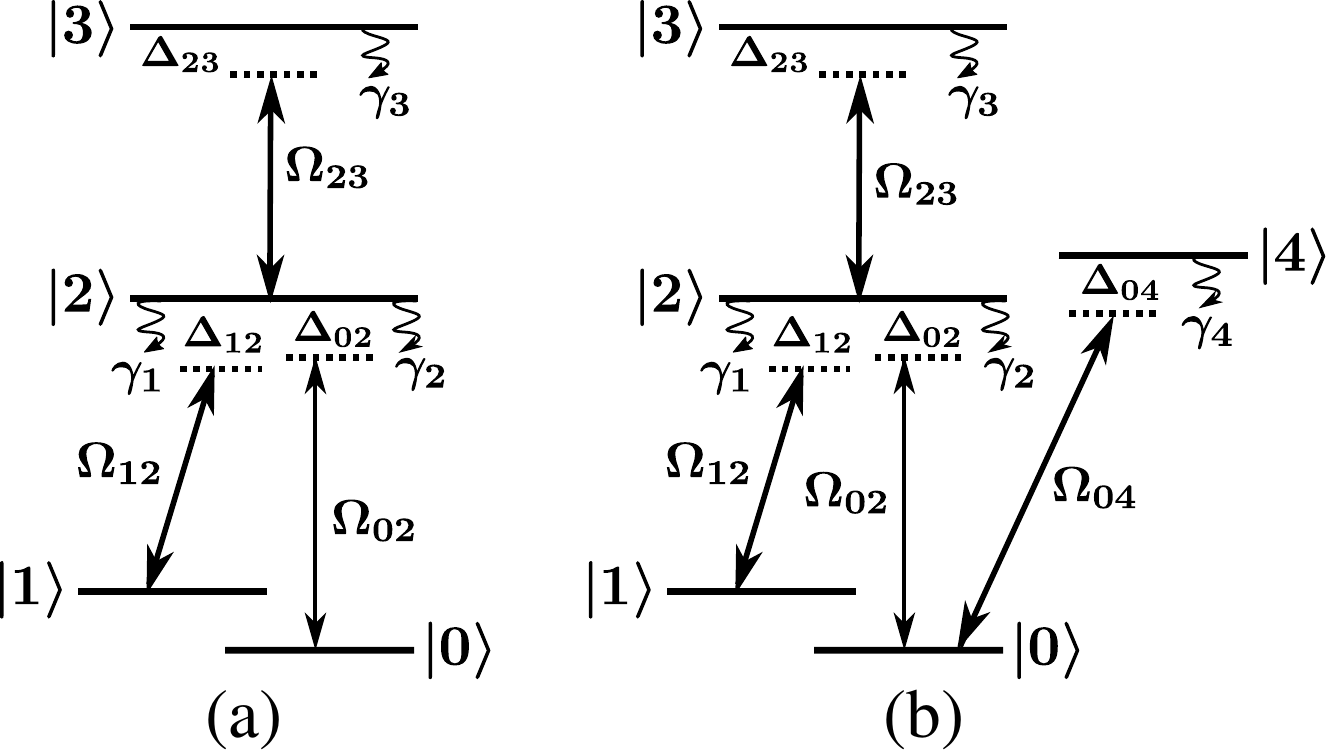}
\caption{\label{figure:level_diagram} The level diagrams of the (a) inverted-Y (IY) atomic system and (b) modified inverted-Y system (described as $IY^+$ system throughout the article). The atomic energy levels are denoted as $\alpha$ with $\alpha \in 0,1,2,3,4$. The coupling strength of individual field connecting states $i$ and $j$ are denoted as $\Omega_{ij}$ and the detuning from corresponding transitions are denoted as $\Delta_{ij}$. }
\end{figure}

\begin{figure}
\includegraphics[width=8.5 cm]{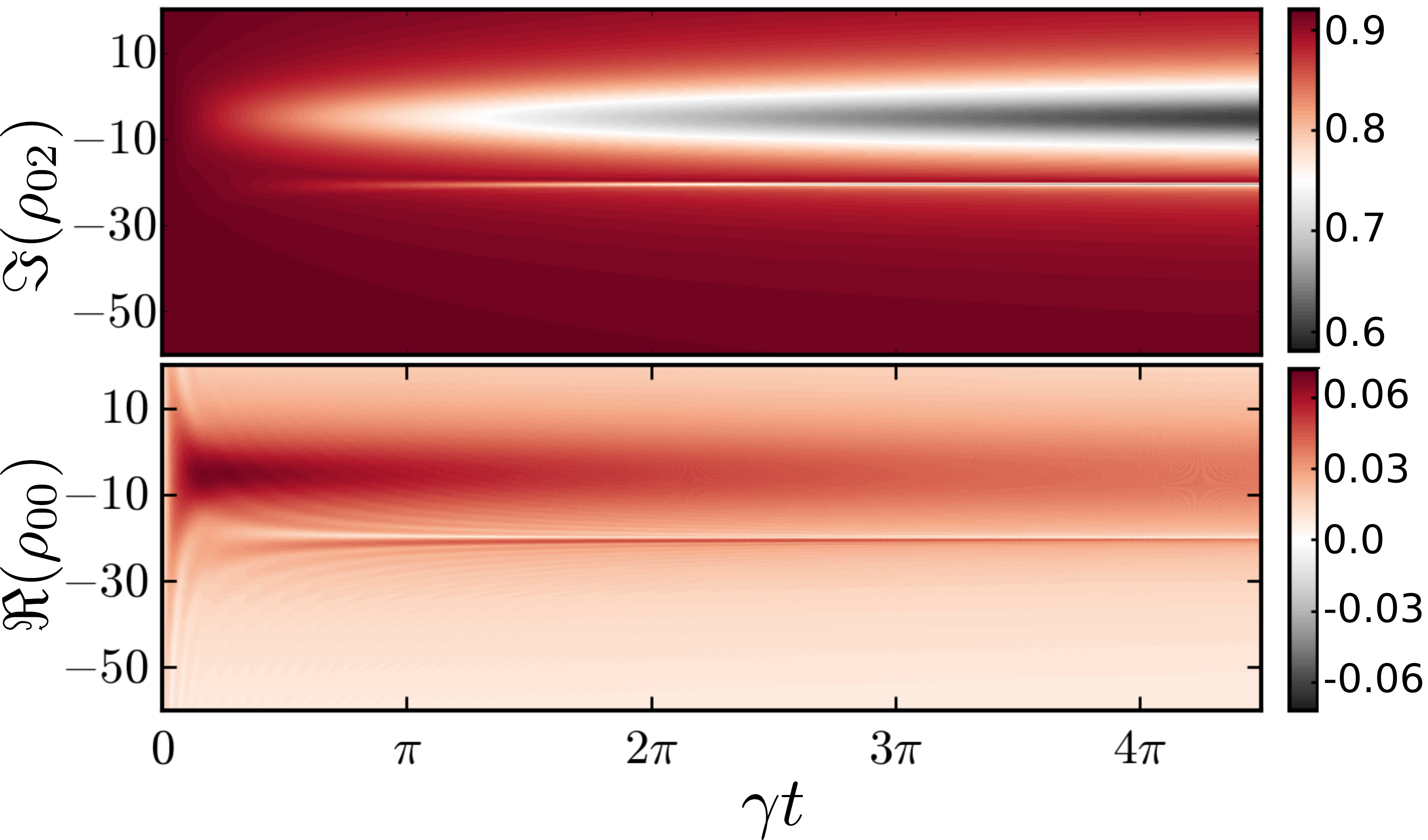}
\caption{\label{figure:transient} (Color online) Amalgamated spectrum of (a) population of the ground state $\Re(\rho_{00})$ and coherence of the probe transition $\Im(\rho_{02})$ as a function of the dimensionless time $\gamma t$. The observed transient characteristics show the atomic system achieving the steady state at $\gamma t\sim\ 4\pi$ with $\Omega_{02}/\gamma=1.0$, $\Omega_{12}/\gamma=5.0$, $\Omega_{23}/\gamma=5.0$, $\Delta_{12}/\gamma=-20$, $\Delta_{23}/\gamma=0$.}
\end{figure}

In order to study the numerical stability of the utilized algorithm, we characterize the transient response of an inverted-Y (IY) system consisting four levels $|\alpha\rangle,\ \alpha\in \{0,1,2,3\}$ along with three electromagnetic fields with coupling strengths $\Omega_{ij}$ connecting states $i$ and $j$ with $i,j \in 0,1,2,3$, as shown in figure  \ref{figure:level_diagram}(a). The population of the ground state $|0\rangle$ and the coherence of the $|0\rangle\longleftrightarrow|2\rangle$ transition of the inverted-Y system as a function of scaled time $\gamma t$ is shown in figure \ref{figure:transient} (a) and (b) respectively. As expected, after initial rapid variations, the steady state is achieved at $\gamma t=\ 4\pi$ and without any external perturbation the atomic system continues to be in the steady state. 

In order to establish the equivalence of the numerical matrix propagation (NMP) method with the RWA method in the low field strength and small detuning regime, the evolution of the density matrix equations has been studied using both the formalisms in the atomic system described above. The density matrix equations of an inverted-Y system (figure. \ref{figure:level_diagram}(a)) can be explicitly written in the rotating wave approximations as,

\begin{eqnarray}
\begin{aligned}
\frac{\partial\rho_{00}}{\partial t}&=2\gamma_2\rho_{22}-2\gamma_0\rho_{00}+i\Omega_{02}(\rho_{02}-\rho_{20})
\nonumber\\
\frac{\partial\rho_{11}}{\partial t}&=2\gamma_1\rho_{22}+2\gamma_0\rho_{00}+i\Omega_{12}(\rho_{12}-\rho_{21})\nonumber\\
\frac{\partial\rho_{22}}{\partial t}&=-2(\gamma_1+\gamma_2)\rho_{22}+2\gamma_3\rho_{33}-i\Omega_{12}(\rho_{12}-\rho_{21})\\&-i\Omega_{02}(\rho_{02}-\rho_{20})+i\Omega_{23}(\rho_{23}-\rho_{32})\nonumber\\
\frac{\partial\rho_{33}}{\partial t}&=-2\gamma_3\rho_{33}-i\Omega_{23}(\rho_{23}-\rho_{32})\nonumber\\
\frac{\partial\rho_{01}}{\partial t}&=-(\gamma_0+i(\Delta_{12}-\Delta_{02}))\rho_{01}+i\Omega_1\rho_{02}-i\Omega_{02}\rho_{21}\nonumber\\
\frac{\partial\rho_{02}}{\partial t}&=-(\gamma_1+\gamma_2-i\Delta_{02})\rho_{02}+i\Omega_{12}\rho_{01}\\&+i\Omega_{23}\rho_{03}+i\Omega_{02}(\rho_{00}-\rho_{22})\nonumber\\
\end{aligned}
\end{eqnarray}

\begin{eqnarray}
\begin{aligned}
\frac{\partial\rho_{03}}{\partial t}&=-(\gamma_0+\gamma_3-i(\Delta_{02}+\Delta_{23}))\rho_{03}+i\Omega_{23}\rho_{02}\\&-i\Omega_{02}\rho_{23}\nonumber\\
\frac{\partial\rho_{12}}{\partial t}&=-(\gamma_1+\gamma_2-i\Delta_{12})\rho_{12}+i\Omega_{02}\rho_{10}\\&+i\Omega_{23}\rho_{13}+i\Omega_{12}(\rho_{11}-\rho_{22})\nonumber\\
\frac{\partial\rho_{13}}{\partial t}&=-(\gamma_3-i(\Delta_{12}+\Delta_{23}))\rho_{13}+i\Omega_{23}\rho_{12}-i\Omega_{12}\rho_{23}\nonumber\\
\frac{\partial\rho_{23}}{\partial t}&=-(\gamma_1+\gamma_2+\gamma_3-i\Delta_{23})\rho_{23}-i\Omega_{02}\rho_{03}-\\&i\Omega_{12}\rho_{13}+i\Omega_{23}(\rho_{22}-\rho_{33})\nonumber\\
\end{aligned}
\end{eqnarray}
and
\begin{eqnarray}
\begin{aligned}
\frac{\partial\rho_{ij}}{\partial t}&=\frac{\partial\rho_{ji}^*}{\partial t},
\end{aligned}
\end{eqnarray}

where $\Delta_{ij}$ is detuning of the applied electromagnetic fields from the corresponding atomic transitions between states $i$ and $j$. $\gamma_1$, $\gamma_2$ and $\gamma_3$ are the decay rates of the excited states $|1\rangle$, $|2\rangle$ and $|3\rangle$ respectively. The total population is conserved by $\sum_{i}\rho_{ii}=1$. The non-radiative relaxation rate of the ground state, i.e. $\gamma_0$, is zero for this closed system.
The steady-state of the above set of equation can be obtained using the matrix method described as,
\begin{equation}
\dot{\rho}=\mathcal{L}\rho=0.
\end{equation}
The eigenvector corresponding to the `zero' eigenvalue of the Liouvillian super-operator ($\mathcal{L}$) correspond to the steady-state value of the density matrix $\rho$.   

\begin{figure}[t]
\includegraphics[width=8.5 cm]{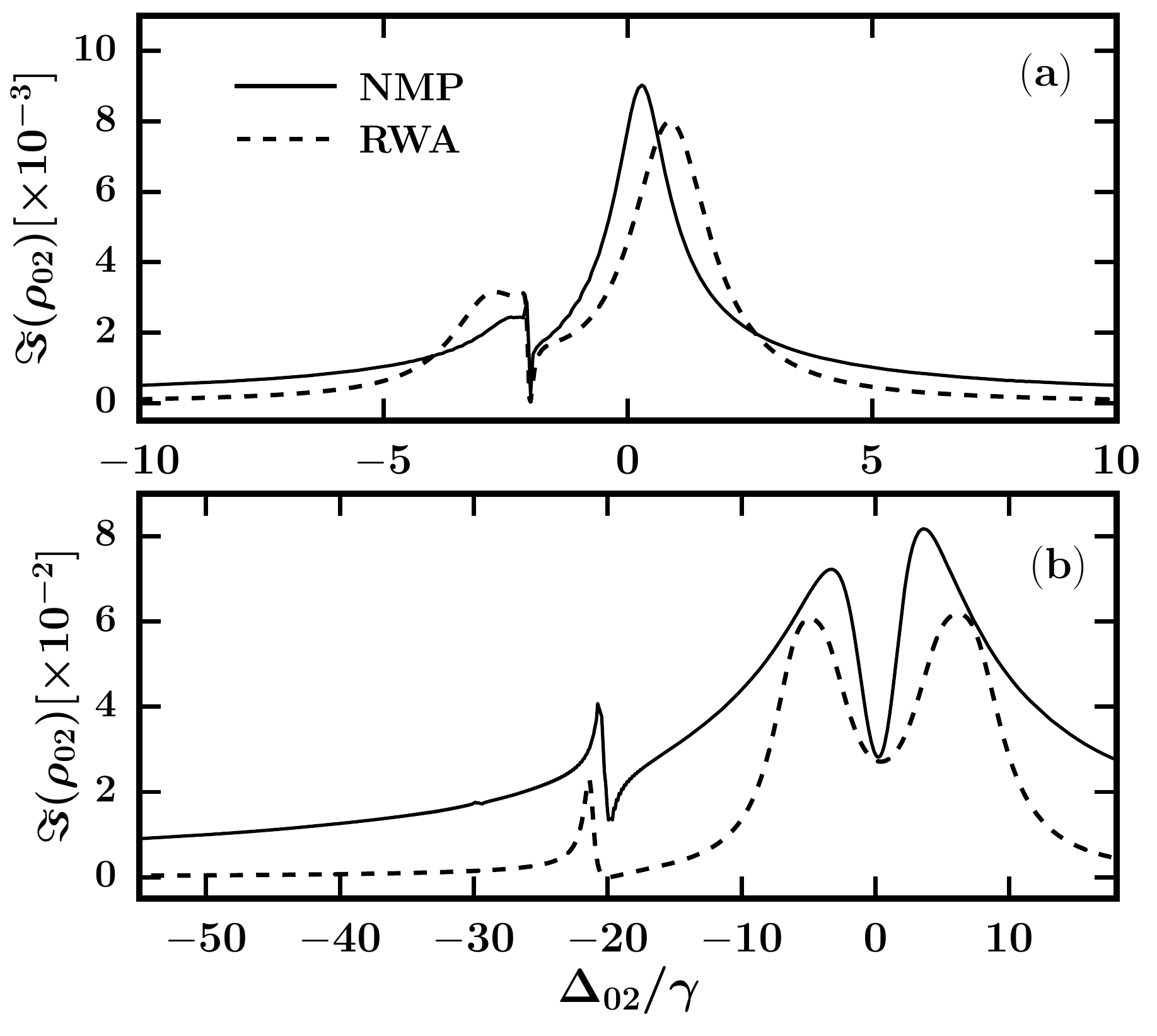}
\caption{\label{figure:rwa_comparison} The probe absorption $\Im(\rho_{02})$ as a function of scaled detuning $\Delta_{02}/\gamma$ with (a) $\Omega_{02}/\gamma=0.01$, $\Omega_{12}/\gamma=0.5$, $\Omega_{23}/\gamma=1.5$, $\Delta_{12}/\gamma=-2$, $\Delta_{23}/\gamma=2$ and (b) $\Omega_{02}/\gamma=1.0$, $\Omega_{12}/\gamma=5.0$, $\Omega_{23}/\gamma=5.0$, $\Delta_{12}/\gamma=-20$, $\Delta_{23}/\gamma=0$. }
\end{figure}

In order to perform the comparison, the steady state value of the density matrix is obtained from both the method for two different coupling strength and detuning regime and the results are shown in figure \ref{figure:rwa_comparison}. Figure \ref{figure:rwa_comparison} (a) correspond to the low coupling strength and low detuning regime with Rabi frequencies $\Omega_{02}/\gamma=0.01$, $\Omega_{12}/\gamma=0.5$ and $\Omega_{23}/\gamma=1.5$ with detuning values $\Delta_{12}/\gamma=-2$ and $\Delta_{23}/\gamma=2$. The additional parameters utilized in the numerical matrix propagation method is $\gamma \delta t=10^{-5}$ propagated upto $\gamma t_{max}=2\pi\times 15$. The equivalence of the RWA (dashed line) and NMP (solid line) methods in low field strength is clearly visible in the graph. The slight shift in the peak of the $\Im(\rho_{02})$ value can be attributed to the counter rotating terms neglected in the RWA method. 

However, in the high field strength and large detuning regime, though the spectral features remains same, the difference between the outcomes of both the methods grow further. In this regime, the field strengths are $\Omega_{02}/\gamma=1.0$, $\Omega_{12}/\gamma=5.0$ and $\Omega_{23}/\gamma=5.0$ with detuning values $\Delta_{12}/\gamma=-20$ and $\Delta_{23}/\gamma=0$. The time sequence parameters of the NMP method were kept same.

\section{Results and discussions}
\label{sec:result}
In the previous section, the stability of the numerical implementation has been presented along with the comparison with the rotating wave approximations. The established validity of the numerical matrix propagation method enables one to employ the same in field strength and frequency regimes commonly not considered in case of RWA.

In this section, a modified inverted-Y system (figure. \ref{figure:level_diagram} (b)) has been studied after incorporating an additional coupling field connecting the ground state to another excited state. This particular choice of an atomic system serves the purpose of integrating all the primitive three level atomic systems namely $\Lambda$, ladder and vee along with an option to study the effect of the merger of two of the basic four level atomic system, \textit{i.e.} the inverted-Y system and N system. The level diagram of the proposed atomic system is shown in figure. \ref{figure:level_diagram} (b) where the unperturbed atomic states are denoted as $|\alpha\rangle$, $\alpha \in \{0,1,2,3,4\}$. The applied electromagnetic fields are described by coupling strength $\Omega_{ij}$, and detuning $\Delta_{ij}$, where subscript $ij$ represents the state $i$ and $j$ through which electromagnetic field couples. The atomic system consist a $\Lambda$ sub-system $|0\rangle\longleftrightarrow|2\rangle\longleftrightarrow|1\rangle$, a ladder sub-system $|0\rangle\longleftrightarrow|2\rangle\longleftrightarrow|3\rangle$ and a vee system $|2\rangle\longleftrightarrow|0\rangle\longleftrightarrow|4\rangle$. The inverted-Y sub-system is formed by the transition paths $|0\rangle\longleftrightarrow|2\rangle\longleftrightarrow|3\rangle$ and $|1\rangle\longleftrightarrow|2\rangle\longleftrightarrow|3\rangle$. Finally the N sub-system is formed via the transitions $|1\rangle\longleftrightarrow|2\rangle\longleftrightarrow|0\rangle\longleftrightarrow|4\rangle$. Assuming the field $\Omega_{02}$ and detuning $\Delta_{02}$ for the transition $|0\rangle\longleftrightarrow|2\rangle$ as a probe field, response of all the above atomic sub-systems can be simultaneously probed. In the proceeding section, the coherence of the density matrix element (\textit{i.e.} $\Im(\rho_{02})$) is analyzed to identify the key spectral features of the $IY^+$ system. If not explicitly described, the field values are the following: $\Omega_{02}/\gamma=1.0$, $\Omega_{12}/\gamma=5.0$, $\Omega_{23}/\gamma=5.0$ and $\Omega_{04}/\gamma=5.0$. The probe beam detuning $\Delta_{02}/\gamma$ has been varied in a wide range $-100\longleftrightarrow 100$ and the relevant part of the spectral response are depicted in the various plots. 

\begin{figure}[b]
\includegraphics[width=8.5 cm]{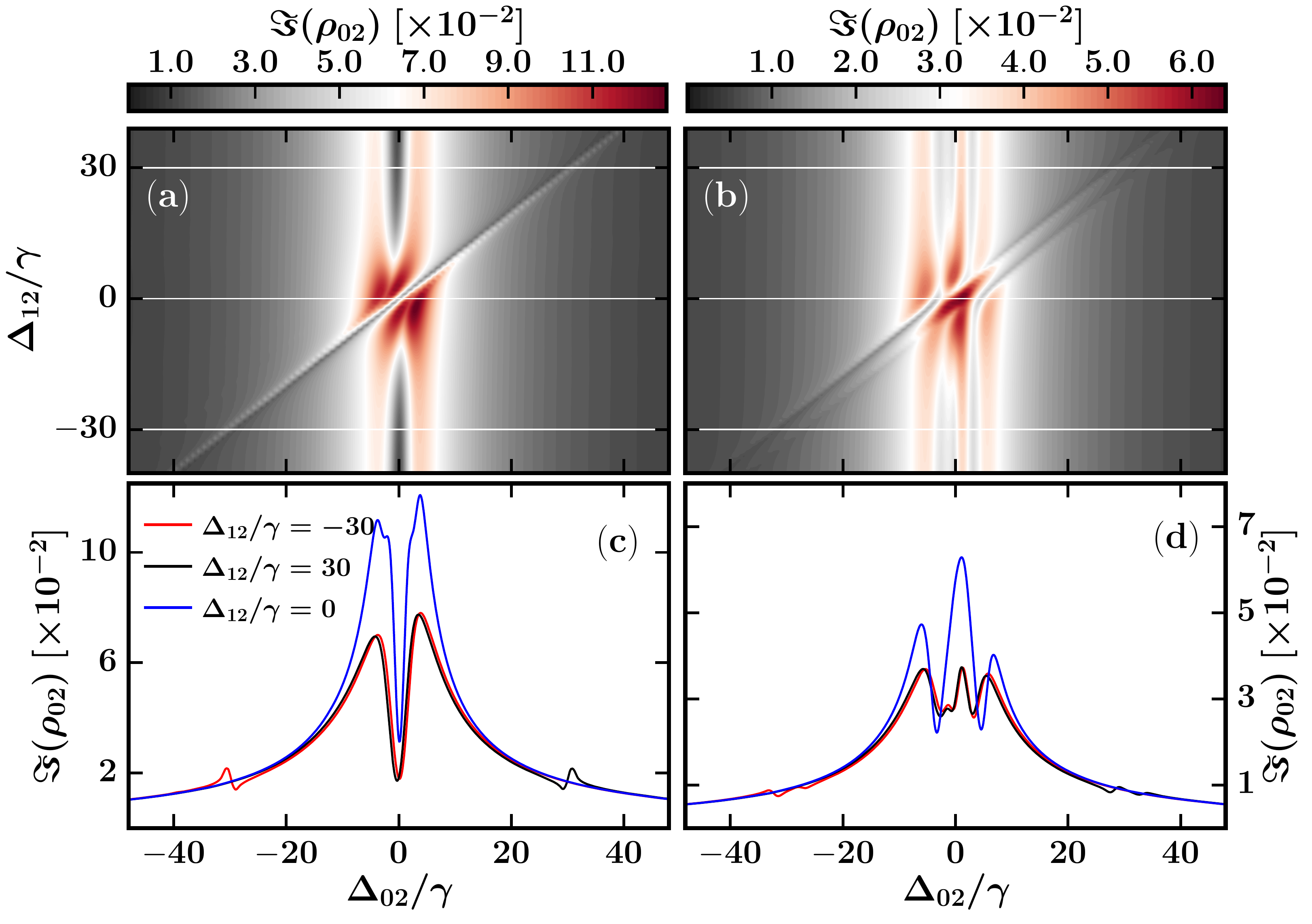}
\caption{\label{figure:d12_variation} (Color online) Amalgamated spectrum (\textit{i.e.} probe absorption versus probe detuning $\Delta_{02}/\gamma$) as a function of the scaled detuning $\Delta_{12}/\gamma$ for (a) inverted-Y system with $\Omega_{02}/\gamma=1.0$, $\Omega_{12}/\gamma=5.0$, $\Omega_{23}/\gamma=5.0$, $\Delta_{23}/\gamma=0$ and (b) $IY^+$ system with parameters $\Omega_{02}/\gamma=1.0$, $\Omega_{12}/\gamma=5.0$, $\Omega_{23}/\gamma=5.0$, $\Omega_{04}/\gamma=5.0$, $\Delta_{23}/\gamma=0$, $\Delta_{04}/\gamma=0$.}
\end{figure}

To begin with, the detuning of the coupling field $\Delta_{12}/ \gamma$ has been varied from $- 40 $ to $ 40 $ for the inverted-Y system as well as the $IY^+$system and the obtained results are shown in figure \ref{figure:d12_variation} . Figure \ref{figure:d12_variation} (a) and (b) show the amalgamated spectrum for inverted-Y and $IY^+$ system respectively and the individual spectrum corresponding to the white lines drawn in plots (a) and (b) are presented in plots  (c) and (d) respectively.
 For inverted-Y system (figure \ref{figure:level_diagram} (a)), as reported in the earlier studies \cite{Qi:2010}, two sharp transparencies (\textit{i.e} EIT) have been obtained at the specific detuning positions where the two photon resonance condition in the $\Lambda$ system $|0\rangle\longleftrightarrow|2\rangle\longleftrightarrow|1\rangle$ and ladder system $|0\rangle\longleftrightarrow|2\rangle\longleftrightarrow|3\rangle$ are satisfied. For both the resonant coupling fields, two EIT dips merge to give a single sharp EIT at zero probe field detuning. The coupling of ground state $|0\rangle$ to excited state $|4\rangle$ via field of strength $\Omega_{04}$ in the $IY^+$ system (figure \ref{figure:level_diagram} (b)) has shown different spectral feature than that of inverted-Y system. At zero detuning of $\Delta_{12}$, figure \ref{figure:d12_variation} (b)) shows higher $\Im(\rho_{02})$  at resonant probe (encoded by red color) implying the existence of an absorption peak rather than a transparency as obtained for the case of inverted-Y system. 

\begin{figure}
\includegraphics[width=7.5 cm]{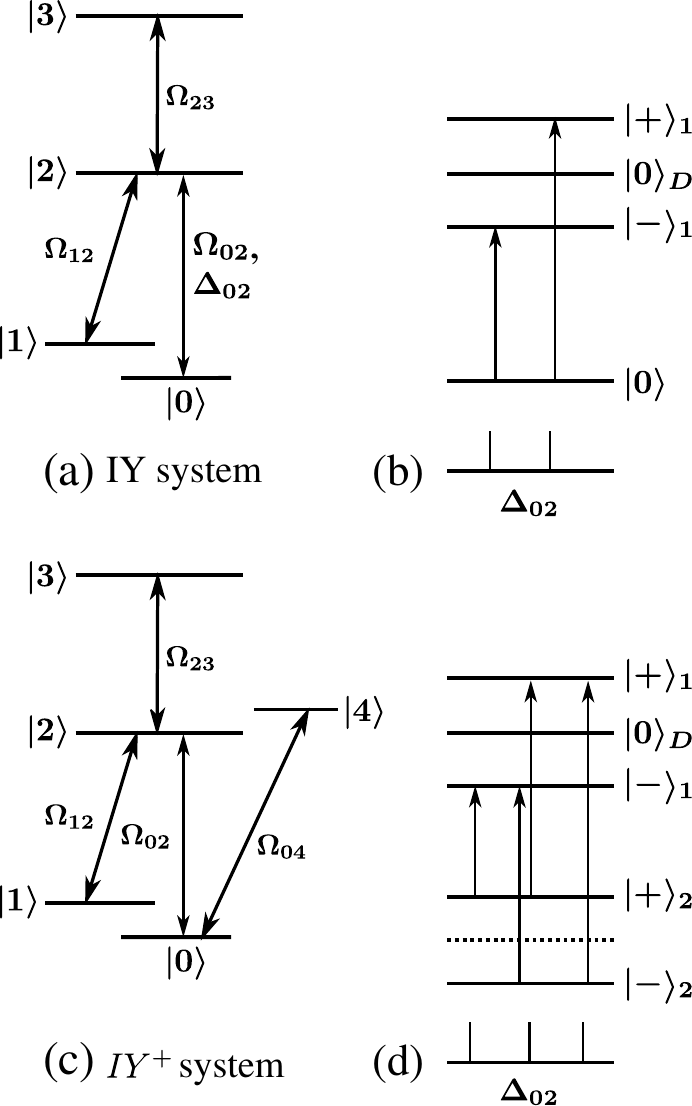}
\caption{\label{figure:dressed} The energy level diagrams of inverted-Y (IY) and $IY^+$ systems. (a) and (c) show bare states with resonant coupling fields, whereas, (b) and (d) show their corresponding dressed states.} 
\end{figure}

A qualitative understanding of such change of spectral behavior from inverted-Y to $IY^+$ system can be obtained using a semi-classical dressed state formalism as explained in the following discussion. In this formalism, the bare states coupled with strong field are transformed into a new atom-photon basis state which is known as dressed state. In the case of resonant driving fields, the dressed states can be determined by diagonalizing the interaction Hamiltonian. The interaction Hamiltonian of the inverted-Y system, after RWA and dipole approximation, can be written as,

\begin{equation}\label{eq:hamiltonian}
H_1=\frac{1}{2} \begin{bmatrix} 
0 &0& &\Omega_{02}& &0  \\
0 &0&  &\Omega_{12}& &0 \\
\Omega_{02} &\Omega_{12}& &0& &\Omega_{23} \\
0 &0&  &\Omega_{23}& &0 \\
\end{bmatrix}.
\end{equation} 

As the strength $\Omega_{02}$ of the probe field is weaker than the strength of all other coupling fields, the probe field can be neglected while evaluating the eigen dressed states corresponding to the Hamiltonian $H_I$ in equation \ref{eq:hamiltonian}. The resulting three eigen-frequencies and their corresponding eigen dressed states are\\

$\lambda_{0,\pm}=0, \pm\frac{1}{2} \sqrt{\Omega_{12}^2+\Omega_{23}^2}$\\

$|0\rangle_D =\frac{\Omega_{23}}{\sqrt{\Omega_{12}^2 + \Omega_{23}^2}} |1\rangle - \frac{\Omega_{12}}{\sqrt{\Omega_{12}^2 + \Omega_{23}^2}} |3\rangle$\\

$|+\rangle_1 =\frac{\Omega_{12}}{\sqrt{2 (\Omega_{12}^2 + \Omega_{23}^2)}} |1\rangle + \frac{1}{\sqrt{2}} |2\rangle + \frac{\Omega_{23}}{\sqrt{2 (\Omega_{12}^2 + \Omega_{23}^2)}} |3\rangle$\\

$|-\rangle_1 =\frac{\Omega_{12}}{\sqrt{2 (\Omega_{12}^2 + \Omega_{23}^2)}} |1\rangle - \frac{1}{\sqrt{2}} |2\rangle + \frac{\Omega_{23}}{\sqrt{2 (\Omega_{12}^2 + \Omega_{23}^2)}} |3\rangle$.\\

The absorption process in this formalism can be understood by determining the transition probabilities between bare state $|0\rangle$ and upper dressed states, which can be expressed as,

\begin{equation}
T_{\alpha \rightarrow \beta} = |\langle \alpha|\vec{d}\cdot \vec{E_0}|\beta \rangle|^2 ,
\end{equation}

where $\alpha$ is ground state, and $\beta \in \{|-\rangle, |0\rangle_D, |+\rangle\}$.

The transition between $|0\rangle$ and $|0\rangle_D$ corresponds to the transition at line center (\textit{i.e.} $\Delta_{02} = 0$) does not exist due to its zero transition probability (\textit{i.e.} $|\langle 0|\vec{d}\cdot \vec{E_0}|0 \rangle_D|^2 = 0$). The other transitions (\textit{i.e.} between $|0\rangle$ and $|\pm\rangle_1$) exhibit non-zero transition probability. As a consequence, the inverted-Y system exhibits EIT at line center along with the two absorption peaks surrounding the EIT. This is schematically shown in figure \ref{figure:dressed} (a) and (b).  

In the $IY^+$ system, the ground state $|0\rangle$ also couples with a state $|4\rangle$ via a strong field $\Omega_{04}$ which results in conversion of the bare ground state into dressed state. Thus, the dressed states for $IY^+$ system include three upper dressed states due to bare states $|1\rangle$, $|2\rangle$, $|3\rangle$ as evaluated for the case of inverted-Y system, and two lower dressed states due to coupling of states $|0\rangle$ and $|4\rangle$ through strong field $\Omega_{04}$. The lower dressed states can be evaluated by diagonalizing two-level interaction Hamiltonian $H_{2}$ after RWA and dipole approximation,\\

$H_{2}=\frac{1}{2}\Omega_{04} |0\rangle \langle 4| + h.c.,$\\

and the eigen-energies and corresponding eigen dressed states are given as, 

$\lambda_2= \pm\frac{1}{2} \Omega_{04}$\\

$|+\rangle_2 =\frac{1}{\sqrt{2}} |0\rangle + \frac{1}{\sqrt{2}} |4\rangle$.\\

$|-\rangle_2 =\frac{1}{\sqrt{2}} |0\rangle - \frac{1}{\sqrt{2}} |4\rangle$.\\

As the upper dressed state $|0\rangle_D$ has zero component of bare state $|2 \rangle$, the transition probabilities $T_{|+\rangle_2 \rightarrow |0\rangle_D}$ and $T_{|-\rangle_2 \rightarrow |0 \rangle_D}$ are zero, while the rest of the transitions have non-zero transition probabilities. This is depicted in figure  \ref{figure:dressed} (d). The calculated location of each transitions and hence the existence of absorption peaks in frequency space are at $\Delta_{02} / \gamma = $ -6.0, -1.0, 1.0, 6.0.

The absorption peaks at $\Delta_{02}/ \gamma =$ -1 and 1 merge with each other and result in a broad single peak. Thus, there are total three peaks observable. The non-zero detuning $\Delta_{12}$ leads to a detuned $\Lambda$ system where the off-resonant two-photon Raman resonance gives rise to an additional absorption peak near zero probe detuning along with the EIT dip at position where two-photon resonance condition of detuned $\Lambda$ system is satisfies \cite{Hemmer:1989}. Consequently, for non-zero $\Delta_{12}$, four absorption peaks at resonance and a EIT at off resonant probe detuning exist. Also, since the eigenvalues and dressed states are actually function of detuning and strengths of all coupling fields,  their variations can result in a change in location of spectral features and their strength as well.

\begin{figure}[b]
\includegraphics[width=8.5 cm]{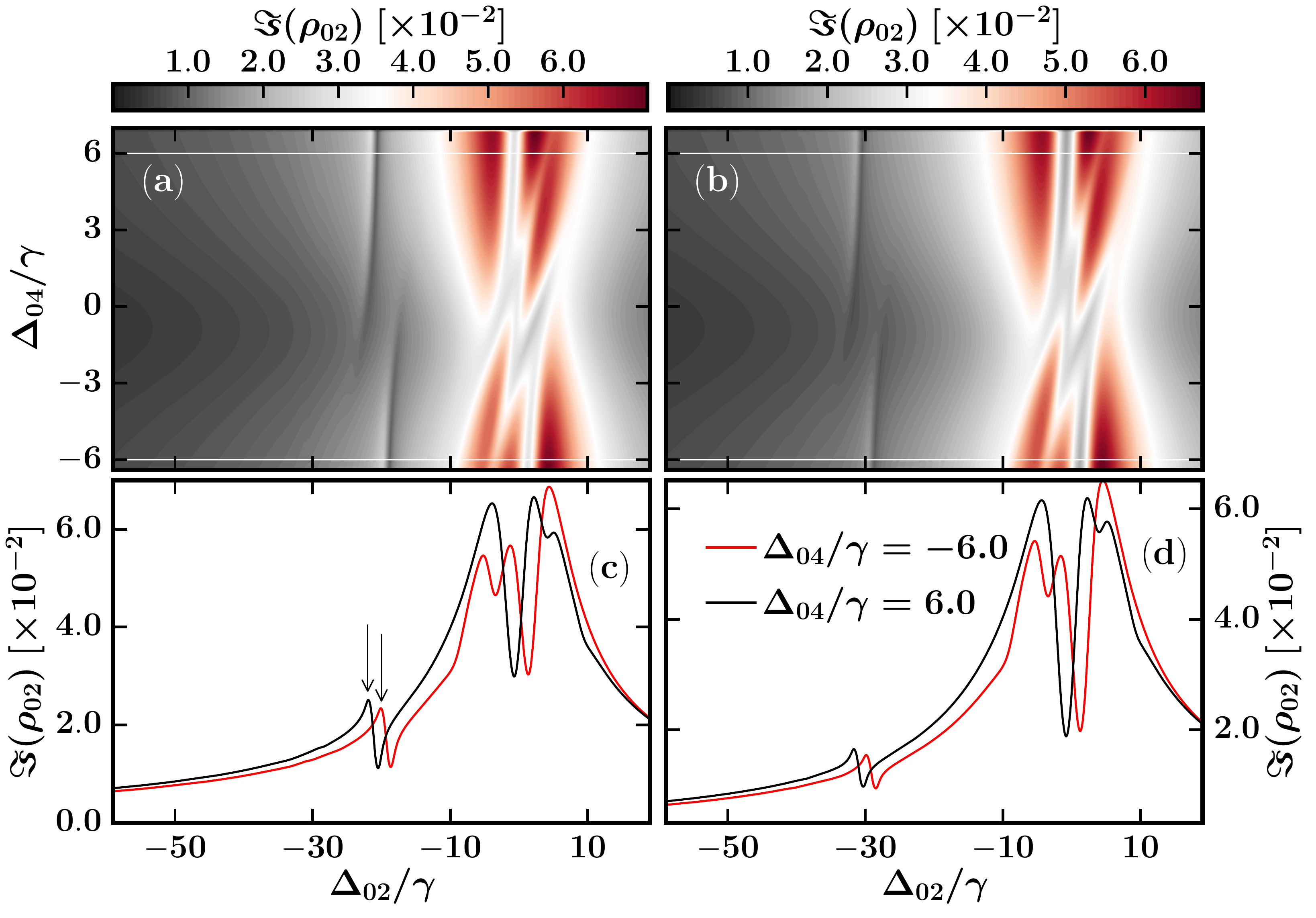}
\caption{\label{figure:d04_variation} (Color online) Amalgamated probe absorption spectrum (\textit{i.e.} probe absorption versus detuning $\Delta_0/\gamma$) as a function of detuning $\Delta_{04}/\gamma$ for (a) $\Delta_{12}/\gamma=-20$ and (b) $\Delta_{12}/\gamma=-30$. The other common parameters are $\Omega_{02}/\gamma=1.0$, $\Omega_{12}/\gamma=5.0$, $\Omega_{23}/\gamma=5.0$, $\Omega_{04}/\gamma=5.0$, $\Delta_{23}/\gamma=0$. The plots (c) and (d) show the probe absorption spectra for values of $\Delta_{04}/\gamma =$ -6.0 and 6.0.}
\end{figure}

The obtained spectral features in above study can further be tailored by varying the detuning $\Delta_{04}$ of another coupling field. For the rest of the studies, the $\Lambda$ system was made far detuned so that its effect can be separated while studying the effect of other systems. The spectra of the probe absorption as a function of detuning $\Delta_{04}$ for $\Delta_{12}/ \gamma =-20$ and $-30$ are shown in left and right column of figure \ref{figure:d04_variation} respectively. Since the electromagnetic field with strength $\Omega_{04}$ directly couples the ground state $|0\rangle$ and an excited state $|4\rangle$, its detuning may affect the coherence between the two ground states $|0\rangle$ and $|1\rangle$. This results in shift in the EIT feature of far-detuned $\Lambda$ system as shown in the figure \ref{figure:d04_variation} (as indicated by an arrow in figure \ref{figure:d04_variation} (c)). Similarly, the resonant spectral features in this figure also show the dependence on the field detuning $\Delta_{04}$. It can be noted here that, in contrast to resonant $\Delta_{04}$ case, where four absorption peaks were observed near the probe resonance frequency,  the resonance features get modified for non-zero $\Delta_{04}$ and  only show three absorption peaks with sharp transparencies between the peaks. The effect of detuned $\Lambda$ system can also be noted by comparing figures \ref{figure:d04_variation} (a) and (b) or  (c) and (d). The detuned $\Lambda$ system affected the spectrum in terms of strength only. This study summarizes that by varying $\Delta_{04}$, the absorption features of probe can be considerably tailored and transparency can be obtained.

\begin{figure}[t]
\includegraphics[width=8.5 cm]{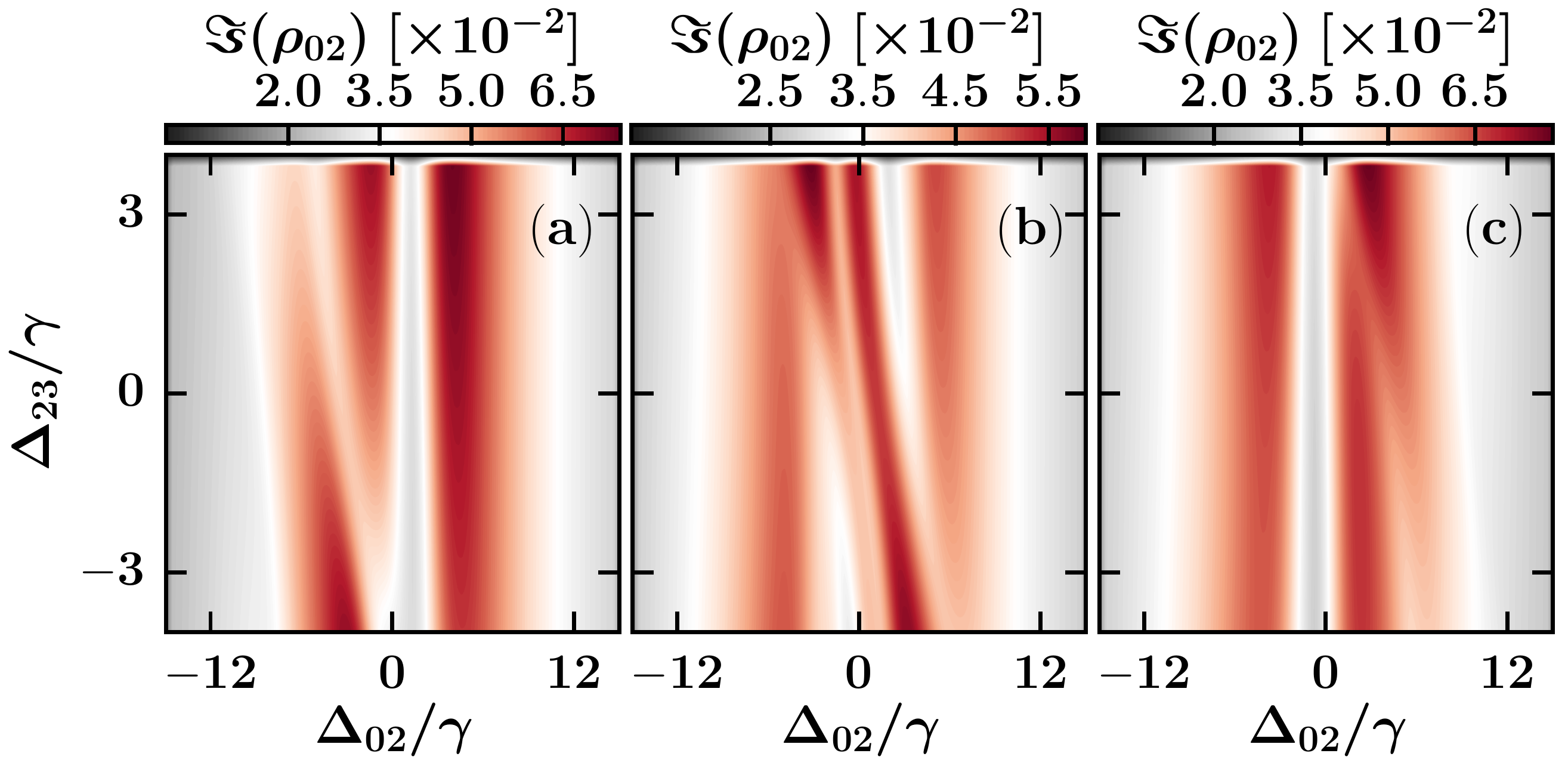}
\caption{\label{figure:d23_variation} (Color online) Amalgamated spectrum ( \textit{i.e.} probe absorption versus the probe detuning $\Delta_{02}/\gamma$) as a function of the scaled detuning $\Delta_{23}/\gamma$ for (a) $\Delta_{04}/\gamma=-6$, (b) $\Delta_{04}/\gamma=0$ and (c) $\Delta_{04}/\gamma=6$ with other parameters $\Omega_{02}/\gamma=1.0$, $\Omega_{12}/\gamma=5.0$, $\Omega_{23}/\gamma=5.0$, $\Omega_{04}/\gamma=5.0$ and $\Delta_{12}/\gamma=-20$.}
\end{figure}

In order to investigate the dependence of the detuned EIT and the resonant spectral features further on other parameters of the externally applied electromagnetic fields, the detuning of another coupling field $\Delta_{23}$ is varied in three different conditions  and the obtained results are shown in figure \ref{figure:d23_variation}. The three different conditions chosen are $\Delta_{04}/ \gamma=-6, 0$ and $6$, as contrasting spectral features have been obtained for these detuning values in previous study. In all the configurations of the detuning $\Delta_{23}$, the detuned EIT shows negligible change and hence is not shown in figure \ref{figure:d23_variation}, whereas the resonant spectra is modified in all the three conditions. This negligible change of EIT may be because of absence of direct decay channel from excited state $|3\rangle$ to ground state $|0\rangle$ resulting in no change in the coherence between ground states $|0\rangle$ and $|1\rangle$.

Figure \ref{figure:d23_variation} (a), (b) and (c) correspond to values of $\Delta_{04}/ \gamma=-6, 0$ and $6$. Previously it has been observed that non-zero $\Delta_{04}$ gives rise to three absorption peaks. Among these three absorption peaks, two of the peaks show variation in its strength with the variation in $\Delta_{23}$, while the third peak is independent of $\Delta_{23}$. The position of these peaks depend on positive or negative values of $\Delta_{04}$. This is clearly visible from figure \ref{figure:d23_variation} (a) and the position of these peaks are inter changed in figure \ref{figure:d23_variation} (c). Along with this, for the case of far detuned $\Delta_{04}$ values \textit{i.e.} $\Delta_{04}/ \gamma= \pm 6$ (figure \ref{figure:d23_variation} (a) and (c)), a single transparency exist at the same probe field frequency for all the values of detuning $\Delta_{23}$. For the resonant case \textit{i.e.} $\Delta_{04}\ / \gamma=\Delta_{23}\ /\gamma=0$ (figure \ref{figure:d23_variation} (b)), there exist four absorption peaks in which the central peak shows a shift in its position with the detuning $\Delta_{23}$. Another central peak shows the change in its strength as well as position with the variation in detuning $\Delta_{23}$ (figure \ref{figure:d23_variation} (b)). In addition to this, a sharp transparency window begins to appear as the detuning $\Delta_{23}$ value is increased in either positive or negative side. Thus, the strength of the absorption peaks can be tuned by applying appropriate detuning $\Delta_{23}$ and to attain a large transparency, $\Delta_{04}$ should be kept non-zero. These spectral characteristics of the system $IY^+$ can be useful for developing optical switching devices.

\begin{figure}[t]
\includegraphics[width=8.5 cm]{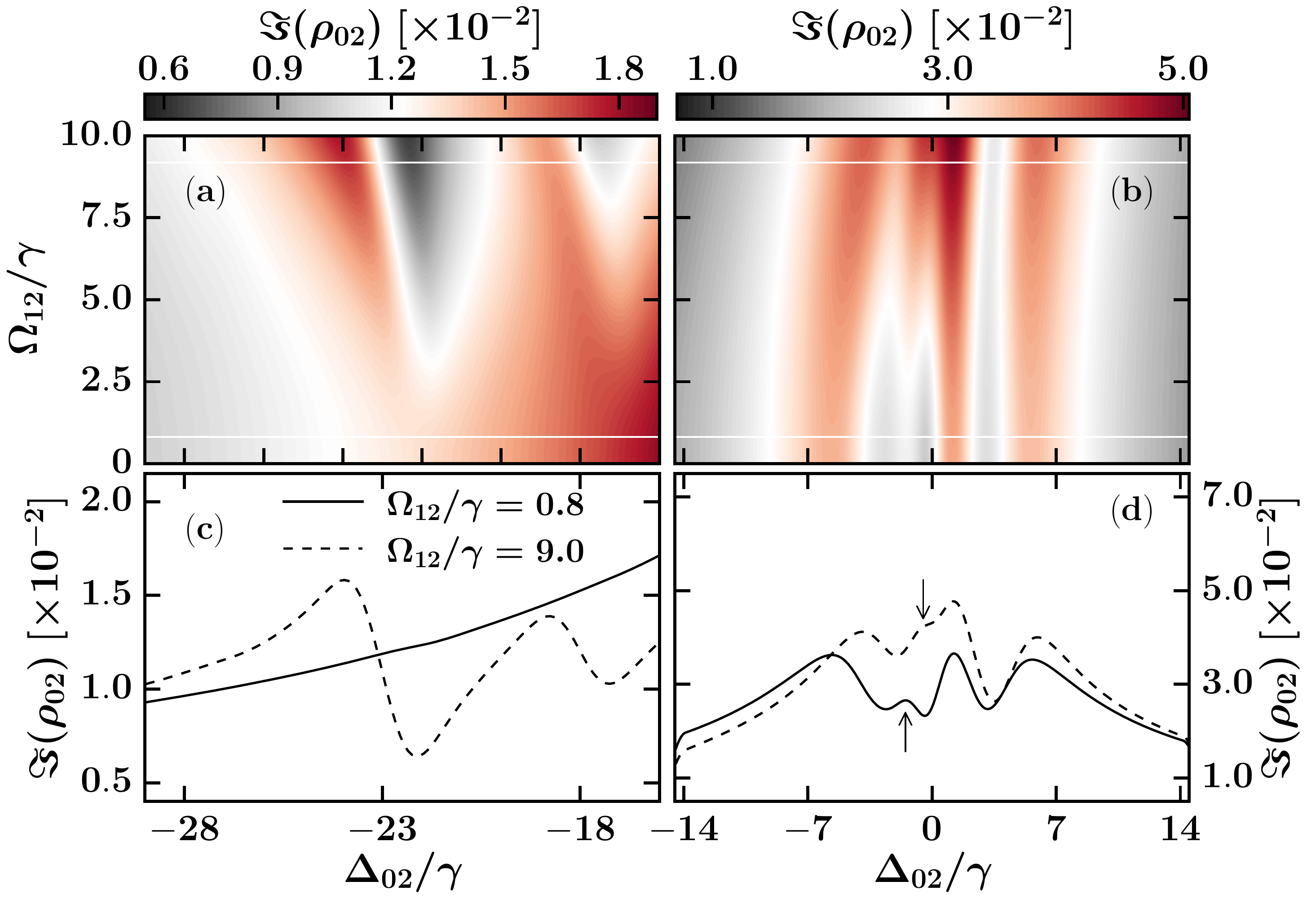}
\caption{\label{figure:O12_variation} (Color online) Amalgamated spectrum (\textit{i.e.} the probe absorption versus probe detuning $\Delta_{02}/\gamma$) for different values of coupling strength $\Omega_{12}/\gamma$. The other parameters are $\Omega_{02}/\gamma=1.0$, $\Omega_{23}/\gamma=5.0$, $\Omega_{04}/\gamma=5.0$, $\Delta_{12}/\gamma=-20$, $\Delta_{23}/\gamma=\Delta_{04}/\gamma=0$. The left and right panels correspond to the negative detuning and near resonant case of probe field respectively. Plots (c) and (d) show individual spectra corresponding to the specific field strengths marked by the white line in plots (a) and (b).}
\end{figure}

Subsequent to the studies on effect of detuning, the spectral features of the probe field have also been studied by varying the strength of all the coupling fields. The variation in probe absorption with the variation in coupling strength $\Omega_{12}$ is shown in figure \ref{figure:O12_variation}. For lower field strength $\Omega_{12}$, the EIT feature (at $\Delta_{02}/\gamma=-20$) corresponding to the detuned $\Lambda$ system ($\Delta_{12}/\gamma=-20$) is not observable. It begins to appear as the field strength $\Omega_{12}$ is increased. The resonant spectral feature also shows dependence on the coupling field strength $\Omega_{12}$. With increase in $\Omega_{12}$ value, the small central absorption peak (shown by an arrow in figure) that appear due to the detuned $\Lambda$ system, merges with the other peaks as shown in figure 8 (b) and (d).

\begin{figure}[t]
\includegraphics[width=8.5 cm]{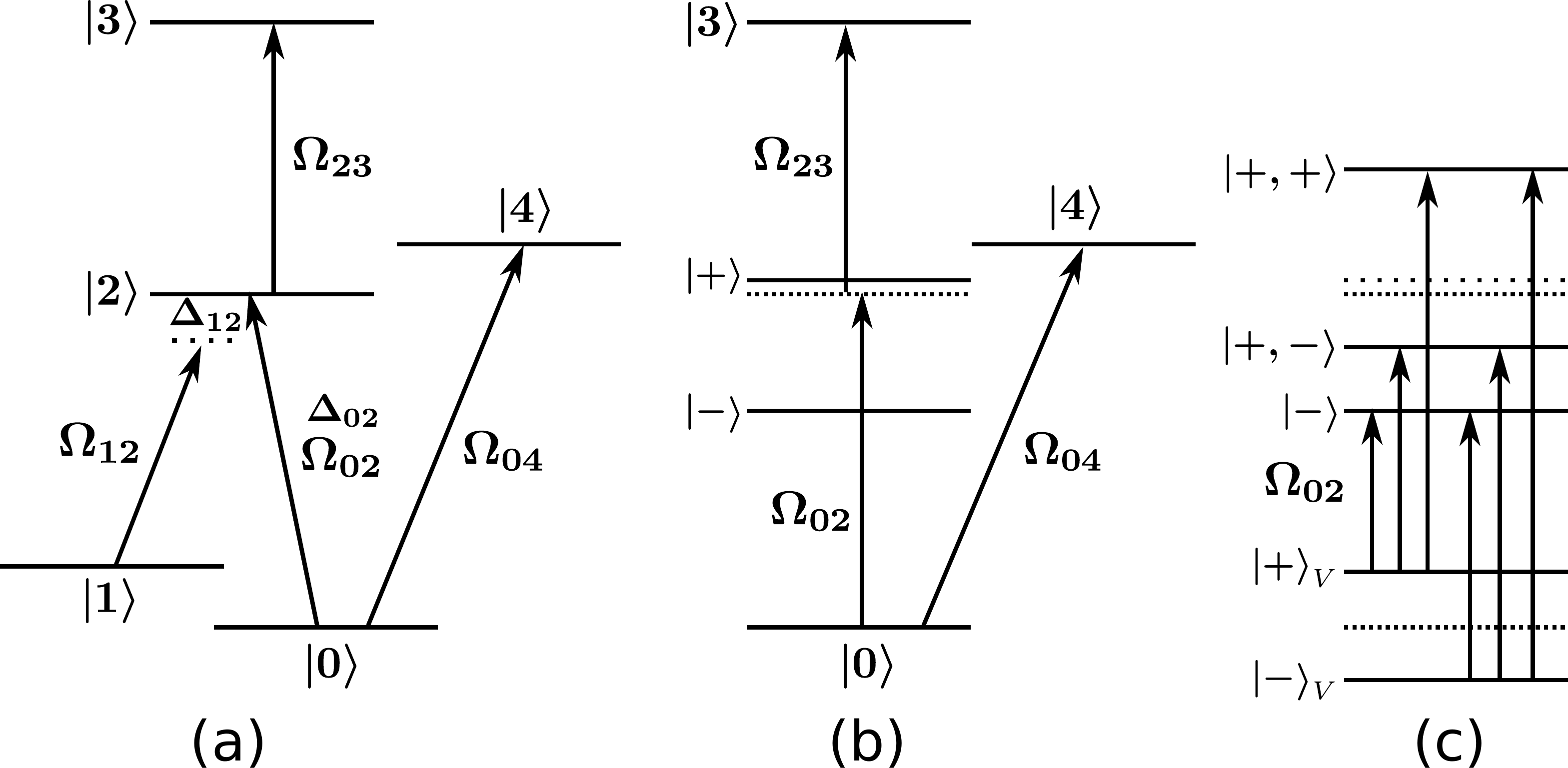}
\caption{\label{figure:O1_dressed}Pictorial representation of doubly dressed approach. (a) Five level $IY^+$-system. (b) The formation of primary dressed states due to strong coupling of $\Omega_{12}$ with states  $|1\rangle$ and $|2\rangle$. (c) The picture representing all possible dressed states with the allowed transitions between them.}
\end{figure}

These obtained results from the NMP method (as discussed above) can be explained using the doubly dressed approach \cite{Yan:2001}, for the case of $\Omega_{ij} > \Omega_{lm}$ where $ij,lm \in \{12,23\}$. In this approach, initially, the bare atomic states coupled with stronger coupling field form the primary dressed states. One of these primary dressed states again gets dressed due to its coupling with another bare state. We considered a particular case for $\Omega_{12}/ \gamma =9, \Omega_{23}/ \gamma=\Omega_{04}/ \gamma = 5, \Omega_{02}/ \gamma = 1, \Delta_{12}/ \gamma=-20, \Delta_{23}/ \gamma=\Delta_{04}/ \gamma = 0$  (figure \ref{figure:O12_variation} (d)). The states $|1\rangle$ and $|2\rangle$ coupled with the strong field of strength $\Omega_{12}$ form primary dressed states $|-\rangle$ and $|+\rangle$ with energies $\frac{1}{2} (\Delta_{12} - \Omega_{12}^{'})$ and $\frac{1}{2} (\Delta_{12} + \Omega_{12}^{'})$ respectively, where $\Omega_{12}^{'}=\sqrt{\Delta_{12}^2 + \Omega_{12}^2}$. The dressed state $|+\rangle$ is again coupled with the bare state $|3\rangle$ through the field $\Omega_{23}$ which creates  doubly dressed states $|+, -\rangle$ and $|+, +\rangle$ with energy $\frac{1}{2} ( \Delta^{'} -\sqrt{\Delta^{'^2} + \Omega_{23}^2})$  and $\frac{1}{2} ( \Delta^{'} +\sqrt{\Delta^{'^2} + \Omega_{23}^2})$  respectively, where $\Delta^{'}=\frac{1}{2}(\Delta_{12} + \Omega_{12}^{'})$.  The formation of these dressed states are depicted in figure \ref{figure:O1_dressed}. The transition of lower dressed states with three upper dressed states via probe field $\Omega_{02}$ give rise to six possible transitions. The calculated location of all these possible transitions in frequency domain are $\Delta_{02}/ \gamma =$ -23.4, -18.4, -3.7, 1.3, 1.4 and 6.4. These calculated locations of transition peaks are in agreement with the obtained results through NMP method as shown in figure \ref{figure:O12_variation} (d).

\begin{figure}[h]
\includegraphics[width=8.5 cm]{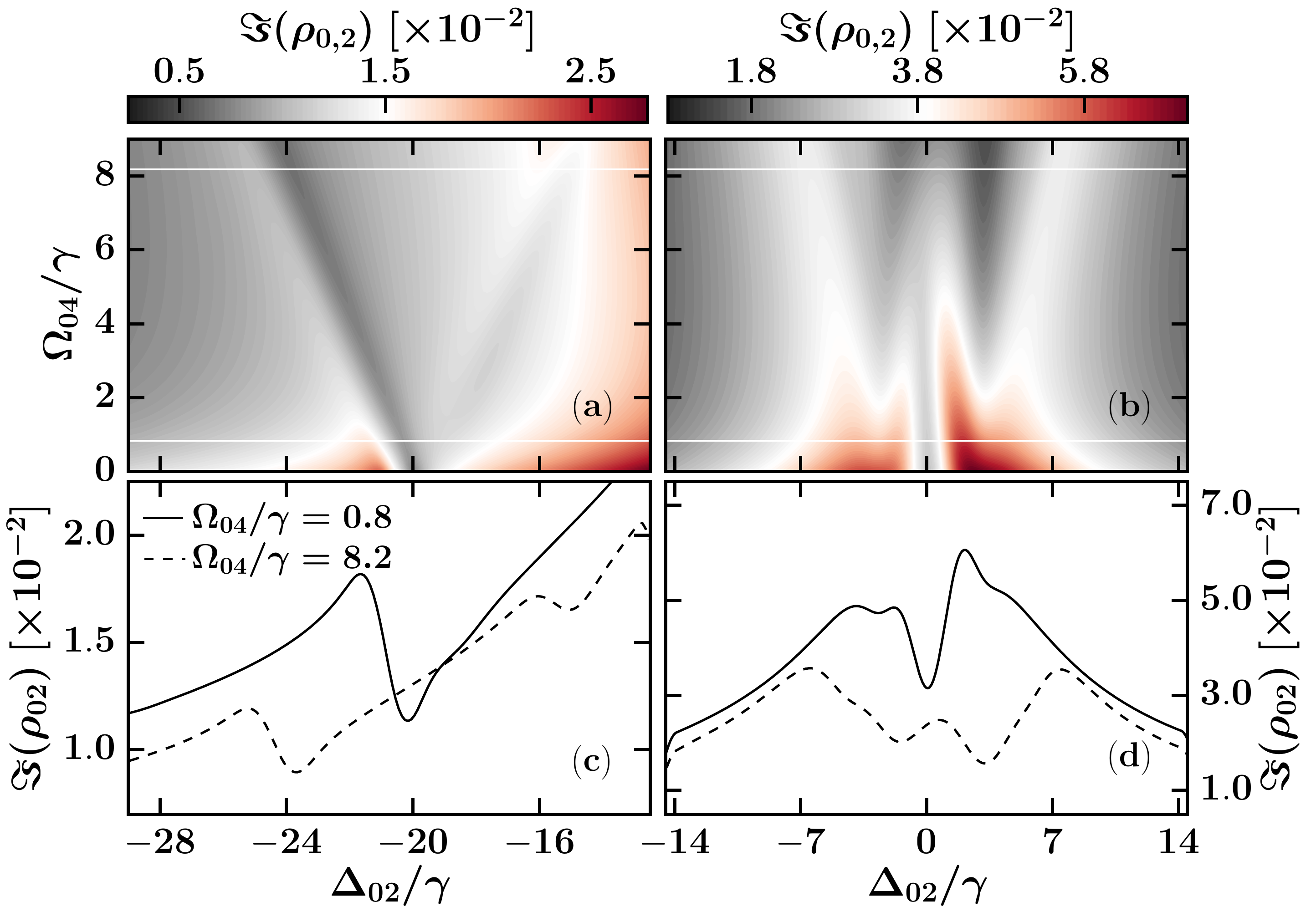}
\caption{\label{figure:O04_variation} (Color online) Amalgamated spectrum (\textit{i.e.} the probe absorption as a function of the probe detuning $\Delta_{02}/\gamma$) for the different values of scaled coupling strength $\Omega_{04}/\gamma$ (a) in far detuned condition and (b) around resonance. Plots (c) and (d) show individual spectra corresponding to the specific field strengths marked by the white lines in (a) and (b). The other parameters are $\Omega_{02}/\gamma=1.0$, $\Omega_{12}/\gamma=5.0$, $\Omega_{23}/\gamma=5.0$, $\Delta_{04}/\gamma=\Delta_{23}/\gamma=0$ and $\Delta_{12}/\gamma=-20$.}
\end{figure}

The study on the effect of variation in coupling strength $\Omega_{04}$ on probe absorption characteristics has also been performed and the corresponding spectrum is plotted in figure \ref{figure:O04_variation}. For a lower value of the coupling strength, \textit{i.e.} $\Omega_{04}=1.8$, one large dispersive EIT feature appears at the detuning position fixed by the $\Delta_{12}$ value. As $\Omega_{04}$ is increased, this large EIT feature splits into two smaller EIT features. With further increase in $\Omega_{04}$, the separation between two EIT peaks increases  (figure \ref{figure:O04_variation} (a) and (c)). This shows that the strength of the coupling field $\Omega_{04}$ modifies the coherence created between two ground states $|0\rangle$ and $|1\rangle$ due to its direct coupling with the state $|0\rangle$ to $|4\rangle$. Using the dressed state approach, this can be attributed to shift in location of the allowed transitions due to increase in the coupling strength $\Omega_{04}$. When we look into resonance spectra of the probe absorption, it is observed that these are also considerably modified as the strength of the coupling field is varied. A large transparency at lower coupling strength $\Omega_{04}$ gets converted into the absorption peak at higher coupling strength $\Omega_{04}$. The observed broad transparency, within which this absorption peak appears, could be a result of depletion in the population from the ground state $|0\rangle$ due to strong coupling $\Omega_{04}$.

\begin{figure}[b]
\includegraphics[width=8.5 cm]{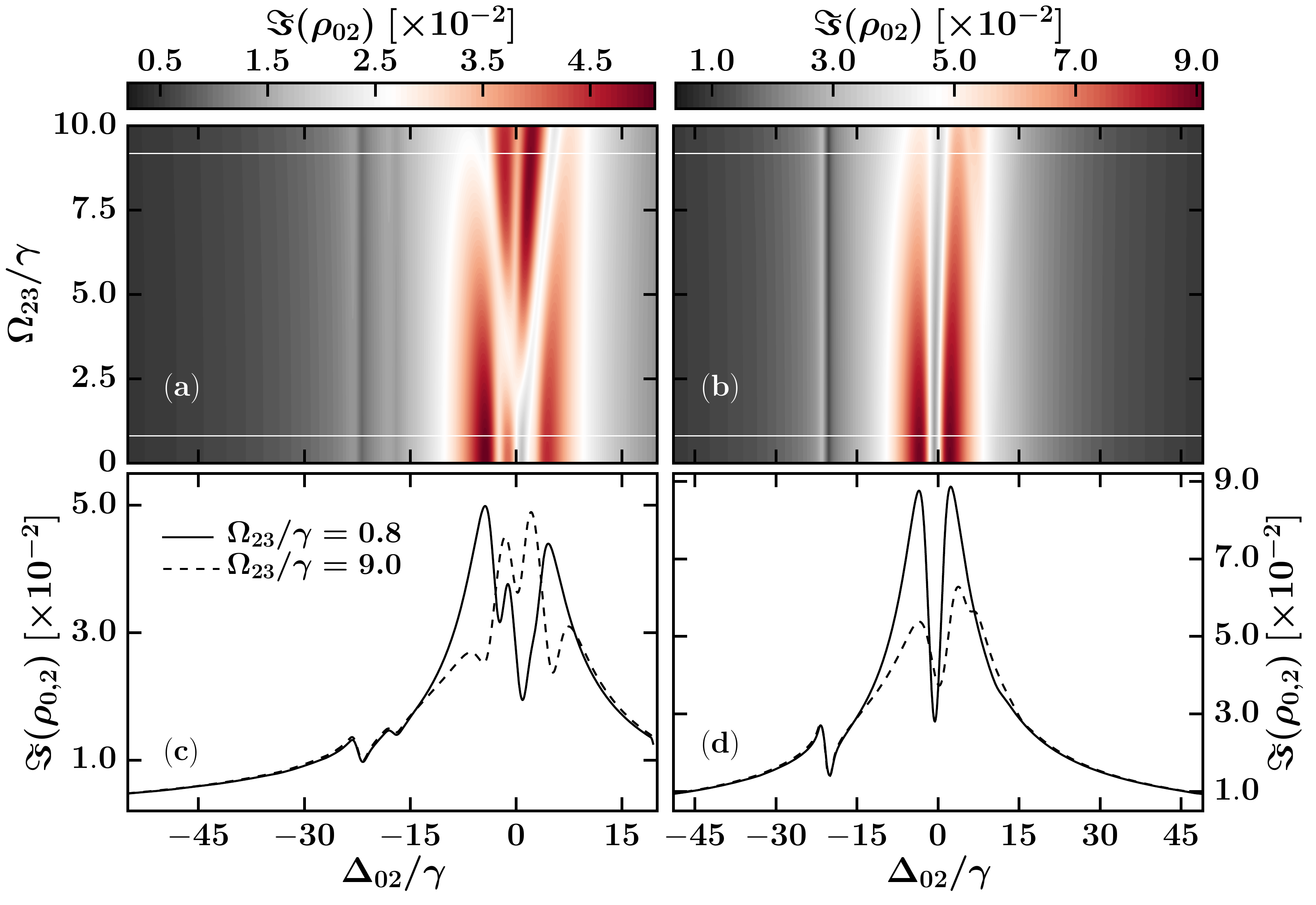}
\caption{\label{figure:O23_variation} (Color online) Amalgamated spectrum (\textit{i.e.} probe absorption as a function of probe detuning $\Delta_{02}/\gamma$) for different values of scaled coupling strength $\Omega_{23}/\gamma$ for (a) $\Delta_{04}/\gamma=0$ and (b) $\Delta_{04}/\gamma=10$ with other parameters $\Omega_{02}/\gamma=1.0$, $\Omega_{12}/\gamma=5.0$, $\Omega_{04}/\gamma=5.0$, $\Delta_{12}/\gamma=-20$ and $\Delta_{23}/\gamma=0$.}
\end{figure} 

\begin{figure}[t]
\includegraphics[width=8.5 cm]{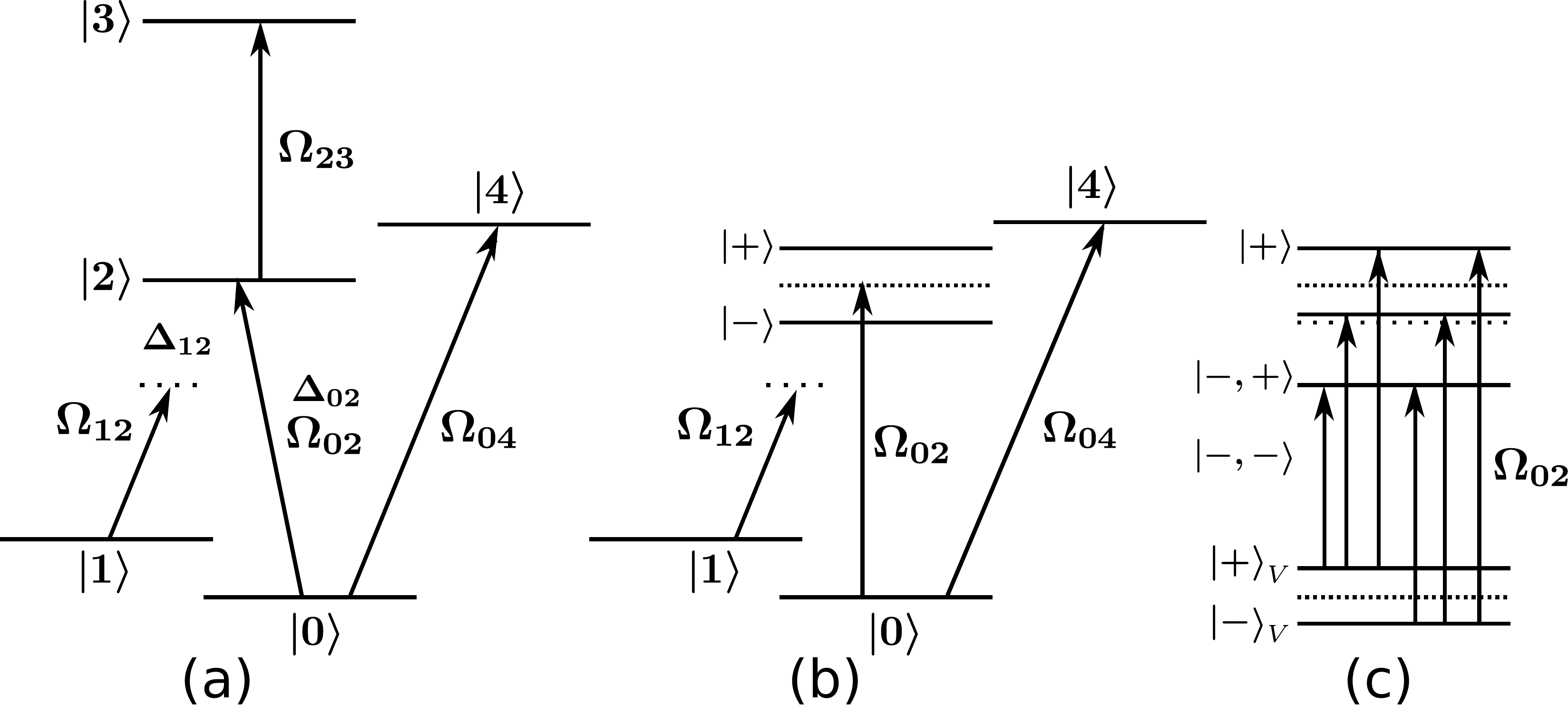}
\caption{\label{figure:O23_dressed}Pictorial representation of doubly dressed approach. (a) Five level $IY^+$ system. (b) The formation of primary dressed states due to strong coupling of $\Omega_{23}$ with states  $|2\rangle$ and $|3\rangle$. (c) The picture representing all possible dressed states with the allowed transitions between them.}
\end{figure}

The effect of variation in coupling strength $\Omega_{23}$ on the probe absorption has been studied for resonant and off resonant detuning values of $\Delta_{04}$. The obtained results are shown in figure \ref{figure:O23_variation}. For the case of resonant condition \textit{i.e.} $\Delta_{04}=0$, one can obtain a coupling strength $\Omega_{23}$ dependent disappearance and reappearance of the absorption peaks in the central region. For a weak coupling strength $\Omega_{23}$, the upper transition between states $|2\rangle$ and $|3\rangle$ acts as a perturbation and the system can be considered as a perturbed N-system. The black continuous curve in figure \ref{figure:O23_variation} (c) shows the corresponding spectrum which is similar to the earlier reported spectrum for the case of N-system \cite{Salloum:2009}. From the figure \ref{figure:O23_variation} (a), it is clear that as strength $\Omega_{23}$ becomes comparable to other coupling strengths, this perturbed N-system becomes equivalent to $IY^+$ system resulting in four absorption peaks as observed earlier (shown by blue curve in figure \ref{figure:d12_variation} (d)). 

For $\Omega_{23}$ greater than the strength of the other coupling fields, the obtained results can be again explained by using the doubly dressed state formalism. The method is same as described earlier. Here, the primary dressed states are created by field $\Omega_{23}$ coupling the bare states $|2\rangle$ and $|3\rangle$. The newly formed dressed states have eigen-energies $\pm \Omega_{23} /2$. The one of the resonantly close primary dressed state (one with energy $-\Omega_{23} /2$) gets doubly dressed due to its interaction with the field $\Omega_{12}$. The energy of these doubly dressed states are $\frac{\Delta_{12} + \Omega_{23}/2}{2} \pm \frac{\sqrt{(\Delta_{12} + \Omega_{23}/2)^2 + \Omega_{12}^2}}{2}$ (see figure \ref{figure:O23_dressed}). The location of all allowed transitions between two lower dressed states and three upper dressed states are expected to be at $\Delta_{02}/ \gamma=$  -22.8, -17.8, -6.6, -1.5, 2.0, 7.0, which is consistent with the results obtained using NMP method (figure \ref{figure:O23_variation} (c)). In case of a far off resonant detuning condition $\Delta_{04}/\gamma=10$, the spectral structure of the central region remains almost independent of the coupling strength $\Omega_{23}$, while amplitude of absorption depends on $\Omega_{23}$.

\section{Conclusion}
\label{sec:conc}
A five-level modified inverted-Y system, \textit{i.e.} $IY^+$ system, comprising of basic three-level sub-systems, \textit{i.e.} $\Lambda$, ladder and vee systems, and basic four-level sub-systems, \textit{i.e.} N and inverted-Y, has been investigated for probe absorption characteristics using a numerical matrix propagation method. The superiority of this method over the well known RWA method has been established by investigating an inverted-Y system within and beyond the validity regime of RWA method. The presence of a strong coupling field connecting the ground state to another state in the $IY^+$ atomic system leads to conversion of resonant probe  transparency (obtained in the inverted-Y system) into absorption. The transparency in $IY^+$ system is recovered when the aforementioned coupling field is kept off-resonant. Apart from this, the coupling field detuning dependent splitting of the transparency and coupling field strength dependent shifting of the transparency and absorption have also been obtained for this $IY^+$ system. The numerically obtained results are also found consistent with the dressed and doubly dressed state formalism. This study shows that the $IY^+$ system can be used to design optical devices for switching and multi-channel optical communication.

\section{ACKNOWLEDGMENTS}
Charu Mishra is grateful for financial support from RRCAT, Indore under HBNI, Mumbai program.


\end{document}